# Convex Optimization of Real Time SoC

L. Yavits, A. Morad, R. Ginosar, U. Weiser

**Abstract**—Convex optimization methods are employed to optimize a real-time (RT) system-on-chip (SoC) under a variety of physical resource-driven constraints, demonstrated on an industry MPEG2 encoder SoC. The power optimization is compared to conventional performance-optimization framework, showing a factor of two and a half saving in power. Convex optimization is shown to be very efficient in a high-level early stage design exploration, guiding computer architects as to the choice of area, voltage, and frequency of the individual components of the Chip Multiprocessor (CMP).

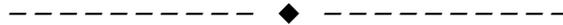

## 1 INTRODUCTION AND RELATED WORK

Optimization became an important part of Chip Multiprocessor (CMP) research in recent years. Typically, performance is the main objective of the CMP optimization [1][4][16][21][24][28]. Recognizing the critical importance of other CMP design aspects such as power consumption, many researchers target optimization of a combination of performance and power, such as energy-delay product [2][3][9]. Some of the optimization studies solve a constrained optimization problem, for example optimizing the performance or energy-delay product under constrained area budget, etc. [2][4][16].

While the performance is a sensible optimization objective in a large variety of CMPs, there is a subclass of CMPs in which it makes less sense. Those are the real time (RT) CMPs, i.e. CMPs which are designed to perform within certain time limits. Examples of such CMPs are video and audio processors and codecs, e.g. [6]. Optimization of such RT CMPs is the subject of the present work. The objective of our optimization framework is the power consumption, while performance (execution time) is addressed as one of the optimization constraints. We extend the MultiAmdahl framework [4][28] by adding the frequency and supply voltage to the optimization problem both as optimization variables and constraints.

Another contribution of this study is the application of convex optimization methodology [22] to power and performance optimization of physical resource-constrained CMP. It has been shown that a wide variety of engineering optimization problems can be reduced to a convex optimization framework [22]. While the majority of prior work solves the optimization problem by analytical techniques (such as Lagrange multipliers) or design exploration by simulation, we formulate the optimization problem as a convex one, and utilize an off-the-shelf solver that finds the global optimum in a very short time. This approach can be efficient for high-level early stage design exploration, guiding computer architects as to the initial choice of area, voltage, and frequency of the individual components of the CMP.

Analytical optimization is a popular optimization tool. Alameldeen [1] used analytical modeling to study the trade-off between the number of CMP cores and cache size. Eliyada et al. [3] optimized frequency and supply voltage for CMP using unconstrained optimization method with an objective to minimize the energy-delay product. Oh at el. [24] suggested optimization of L1 and L2 cache sizes by partitioning of a constrained cache area among them, and evaluating the effect of such partition on the resulting performance of the CMP. Rotem et al. [9] developed tools for unconstrained optimization of power consumption in CMP. Cassidy et al. [2] studied area-constrained CMP optimization and optimal area allocation among core and cache using an energy-delay objective function. Present authors [16] researched optimal allocation of constrained resources among a CMP cache hierarchy levels. Zaidenberg et al. [28] introduced MultiAmdahl, a resource constrained optimization framework for optimal resource allocation in CMP. Morad et al. [4] researched optimal accelerator selection and area division among them in heterogeneous CMP.

A number of simulation based optimization studies have also been conducted. Isci et al. [7] developed a global CMP power management policy, optimizing the performance at a given power budget. Heo et al. [23] designed a framework of reducing the power density and peak temperature by moving computation between multiple replicated units. Zhao et al. [18] developed a constraint-aware analysis methodology that uses chip area and bandwidth as constraints. Huh et al. [12] evaluated cores of different sizes and concluded that lack of bandwidth scaling will promote the use of larger cores.

This research uses analytical modeling. We utilize the performance, execution delay and power models developed by Wentzlaff et al. [8], Hardavellas et al. [21], Loh [10], Butts et al. [11] and Yavits et al. [16][17]. The analysis is applied to an industrial SoC [6].

The rest of this paper is organized as follows. Section 2 presents the experimental setup for the case study. Section 3 describes the analytical models used in optimization. Section 4 details the optimization framework. Section 5 offers

---

- *Leonid Yavits, E-mail: yavits@tx.technion.ac.il.*
- *Amir Morad, E-mail: amirm@tx.technion.ac.il.*
- *Ran Ginosar, E-mail: ran@ee.technion.ac.il.*
- *Uri Weiser, E-mail: uri.weiser@ee.technion.ac.il*

*Authors are with the Department of Electrical Engineering, Technion-Israel Institute of Technology, Haifa 32000, Israel.*

## 2 EXPERIMENTAL SETUP

We apply optimization to an industry RT MPEG2 encoder System-On-Chip (SoC) [6][15] (Fig. 1). This MPEG2 encoder is a heterogeneous CMP, designated for TV broadcast applications, performing real time video, audio and transport stream encoding. The SoC [15] encodes 30 Standard Definition (SD) video frames per second, 39 audio frames (comprising one second of audio), and multiplexes video and audio bit streams into a MPEG2 transport stream. The MPEG2 encoder SoC (Fig. 1) consists of the following processing units (accelerators): Video Input Processor (VIP), Digital Signal Processor (DSP), Motion Estimator (ME), BitStream Processor (BSP), Transport Multiplexor (MUX), Audio Encoder (AUD), Central Controller (CCTR) and DRAM Controller (MCTR). All these units have different architectures and instruction sets.

The operation in the RT MPEG2 encoder SoC of Fig. 1 is pipelined, such that each video processing unit (VIP, DSP, ME, BSP) processes one video macroblock (a 16x16 pixel segment of a video frame) at a time. The AUD must be synchronized to the video processing units, to process the relevant portions of audio frames in each video frame period (approximately 30ms per SD frame). MUX is more loosely synchronized to the video and audio processing pipelines but it also must maintain strict real time and buffer size requirements. Hence the execution delay of all internal processing units of the MPEG2 encoder is strictly constrained by the RT requirements. The difference between this SoC and a non-real time CMP is that there is no need to minimize the execution delay beyond that constraint, since there is no need to encode broadcast video and audio faster than their "natural" frame rate.

In Section 4 below we derive the execution delay constraints, the area and power figures for each processing unit from the MPEG2 encoder [15] simulation.

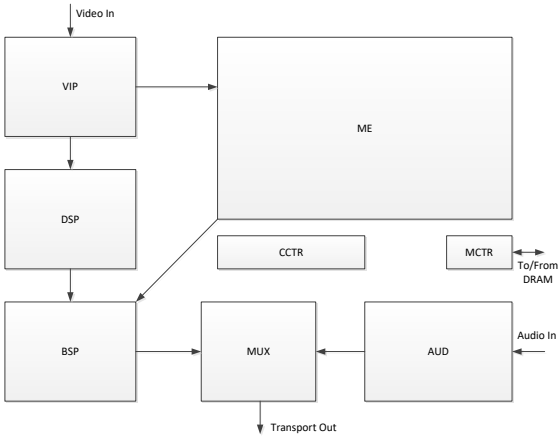

Fig. 1. MPEG2 Encoder Architecture [15]

## 3 ANALYTICAL MODEL

The execution delay of a processing unit (accelerator) can be presented as

$$T_j = t_j f_j(A_j) \frac{F_{REF}}{F_j}, j = 1, \dots, n \quad (1)$$

where $n$ is the number of processing units, $t_j$ is a baseline (un-accelerated) delay, $A_j$ is $j^{th}$ processing unit area, $f_j(A_j)$ is the inverse speedup of the $j^{th}$ accelerator as a function of its area, relative to a baseline implementation of the accelerator with area $A_{MINj}$ and operating frequency $F_{REF}$; $F_j$ is the operating frequency of the $j^{th}$ accelerator.

Following the methodology established in [20], [28] and [4], we express the inverse speedup of a processing unit as a power law of its area, as follows:

$$f_j(A_j) \propto A_j^{-\mu}, \quad 0.3 \leq \mu < 1 \quad (2)$$

The exponent $\mu$ spans from 0.3 for units which are difficult to accelerate by allocating additional area (such as the CCTR), through 0.5 (for a general purpose processor such as the AUD) to 0.95 (for massively parallel fine grain array architecture such as the ME). One possible technique of estimating $\mu$ is presented in Fig. 2(a) for the ME unit and in Fig. 2(b) for the DSP unit respectively. Fig. 2 plots the [area-inverse speedup] points corresponding to a number of potential configurations of each unit. The $\mu$ value is obtained by least square interpolation, also shown in the figure.

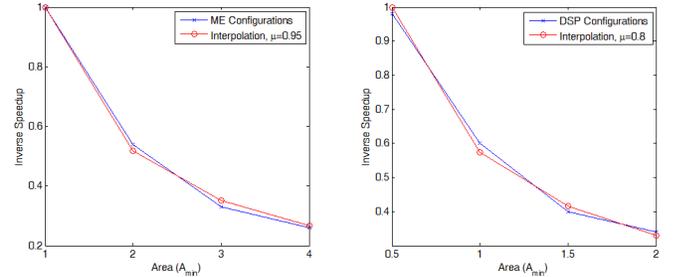

Fig. 2. Example of $\mu$ estimation for (a) ME and (b) DSP

The relation between the operating frequency and supply voltage depends on the voltage level. We assume the following power law operating frequency - supply voltage dependency:

$$F_j \propto V_j^\alpha, \quad 1 \leq \alpha \leq 3 \quad (3)$$

where $\alpha$ is 1 for a normal operating range ($0.8V \leq V_j \leq 1.2V$) [3][9], but can be as much as 3 for near-threshold voltage levels [14].

The dynamic power consumption of a unit $P_j$ can be presented as follows:

$$P_j \propto A_j F_j V_j^2, \quad j = 1, \dots, n \quad (4)$$

The static (leakage) power of a unit $P_{Lj}$ can be presented as follows [11]:

$$P_{Lj} \propto A_j V_j, \quad j = 1, \dots, n \quad (5)$$

## 4 OPTIMIZATION FRAMEWORK

In this section we define the convex optimization, formulate the optimization problem, and present the results of optimization using the RT MPEG2 encoder case study.

## 4.1 Convex Optimization

A mathematical optimization problem has the form:

*Minimize $f(x)$*
*Subject to $g_i(x) \leq L_i, i = 1, \ldots, m$*

where $x = (x_1, \ldots, x_n)$ is the optimization variable of the problem, $f(x)$ is the objective function, $g_i(x)$ are the constraint functions and $L_i$ are the limits for the constraints. Convex optimization problem is one in which the objective and the constraint functions are convex, i.e. they satisfy the inequality $f(\alpha x + \beta y) \leq \alpha f(x) + \beta f(y)$ [22].

## 4.2 Problem Formulation

The performance optimization problem (minimization of the execution delay) under constrained area, supply voltage and operating frequency can be presented as follows:

$$\text{Minimize} \quad \max\left(T_j = t_j f_j(A_j)\frac{F_{REF}}{F_j}\right)$$

$$\text{Subject to} \quad \sum A_j \leq A_{TOT}$$
$$F_j \propto V_j^\alpha$$
$$A_{MINj} \leq A_j \leq A_{MAXj}$$
$$F_{min} \leq F_j \leq F_{max} \quad (6)$$
$$V_{min} \leq V_j \leq V_{max}$$

$$x = \{A_1, \ldots, A_n, F_1, \ldots F_n, V_1, \ldots, V_n\}, \; j = 1, \ldots, n$$

If we substitute $F_j = F_{REF}$ in (6), our optimization problem reverts to MultiAmdahl [28].

In non-RT CMP, maximizing performance is a legitimate optimization goal. However, in RT CMP, maximizing the performance may lead to performance "slack", so that execution completes faster than required by the RT constraint. This typically comes at the cost of excessive power consumption. As we show in this study, such power redundancy may be disproportionate to the execution delay gain which it enables, so that the total execution energy is higher.

We suggest using the power consumption as an objective function, and the execution delay as a constraint. By optimizing the power, we also optimize the area, since both dynamic and static power consumptions are proportional to the die area. The optimization problem can be written as follows:

$$\text{Minimize} \sum_{j=1}^{n}(P_j + P_{Lj})$$

$$\text{Subject to} \quad \sum A_j \leq A_{TOT}$$
$$t_j f_j(A_j)\frac{F_{REF}}{F_j} \leq T_{MAXj}$$
$$F_j \propto V_j^\alpha$$
$$A_{MINj} \leq A_j \leq A_{MAXj} \quad (7)$$
$$F_{min} \leq F_j \leq F_{max}$$
$$V_{min} \leq V_j \leq V_{max}$$

$$x = \{A_1, \ldots, A_n, F_1, \ldots F_n, V_1, \ldots, V_n\}, \; j = 1, \ldots, n$$

where $T_{MAXj}$ is the per-unit RT constraint. The limit values for MPEG2 encoder [6][15] are summarized in TABLE 1.

TABLE 1
LIMIT VALUES FOR MPEG2 ENCODER UNITS [6][15]

| Variable | Min Value | Max Value |
|---|---|---|
| $F_j$ | $100 MHz \; \forall j$ | $2 GHz \; \forall j$ |
| $V_j$ | $0.8V \; \forall j$ | $1.2V \; \forall j$ |
| $A_j$ | $\{.05 \; .2 \; .4 \; .5 \; .1 \; .5 \; .5 \; 1\} mm^2$ | $\{.1 \; .3 \; .6 \; .8 \; .2 \; 2 \; 1 \; 4\} mm^2$ |

*The order of units is {MCTR BSP MUX AUD CCTR DSP VIP ME}*

Since the subject of our optimization framework is an application specific SoC, varying the unit area in the $A_{MINj} - A_{MAXj}$ range results mainly in duplicating or removing computational elements and alike. It does not involve changing the microarchitecture. Therefore the area change is not expected to affect the operating frequency, which allows an independent optimization of the latter.

We use cvx MATLAB solver [19] to solve the optimization problems (6) and (7). Since the objective functions in both problems are not convex, we convert them into a convex representation by logarithmization before using the solver. Optimization runs are performed on Intel® Core2™ Q8400 CPU with operating frequency of 2.67GHz and 8GB RAM, and take approximately 10 sec per run.

## 4.3 Case Study

Fig. 3 summarizes the results of convex optimization applied to the RT MPEG2 encoder [15] of an industrial SoC [6]. Fig. 3(a) shows the optimal area per unit for optimization problems (6) and (7); Fig. 3(b) presents the optimal frequencies per unit; Fig. 3(c) shows the optimal voltage supply values per unit; and Fig. 3(d) presents the optimal power consumptions per unit.

When optimizing for power, the area allocation is lower for all units (compared to execution time optimization) except for VIP and ME. The latter is counterintuitive, since power is proportional to area and hence common sense requires the area to be minimal in a power-optimal design. The reason for the exception is the higher activity and degree of parallelism of VIP and ME ($\mu$ is 0.9 and 0.95 accordingly) comparing to the rest of units (with $\mu$ under 0.8).

The frequencies are generally lower by 20%-25% in the power-optimal design.

The supply voltages differ quite significantly, with power-optimal design taking the lower end of the scale (below 0.9V) and time-optimal design spanning the upper end of the scale (around 1.2V). The reason is the quadratic dependency of dynamic power on the supply voltage.

The difference in power consumption is most apparent in large massively parallel units. It reaches ~2.6 times for ME and VIP while slightly decreasing in predominantly sequential, less power hungry units.

The optimization results exhibit little dependence on the value of $\alpha$ ((3)). The main difference between $\alpha = 3$ (Fig. 3) and $\alpha = 1$ is that the supply voltage in power-optimal design drops to the minimum (0.8V) for all units, while for $\alpha = 3$, it varies in the range of 0.8V to 0.9V.

When minimizing the power consumption, the execution delay of each processing unit equals its RT constraint (30 ms in the RT MPEG2 encoder case study). When minimizing the max $T_j$, the worst execution delay reaches 80% of the RT constraint. On the other hand, the power consumption of the performance-optimal MPEG2 encoder is 2.5 times the power consumption of the power-optimal design on average. Hence the total execution energy of the power-optimal design is half the execution energy of the execution time-optimal design.

## 5 CONCLUSIONS

Performance is a typical objective in CMP optimization. However there is a subclass of CMPs in which optimizing the performance makes less sense. Those are the real time (RT) CMPs, i.e. CMPs which are designed to perform within certain time limits. Optimization of such RT CMPs is the subject of the present work. The objective of our optimization framework is the power consumption, while performance (execution time) is addressed as one of the optimization constraints.

We apply a convex optimization framework to optimize the MPEG2 encoder [15] of an industrial SoC [6], targeting the power optimization under the RT constraint. We compare it to a conventionally performance-optimized MPEG2 encoder. We find that the performance-optimal MPEG2 encoder reaches better than needed performance at a cost of higher power. In contrast, a power-optimal MPEG2 encoder reaches much lower power consumption, while maintaining the RT execution requirements.

We find that convex optimization is a powerful tool for optimizing complex architectures with a variety of physical constraints. It allows computer architects to obtain the globally-optimal results while considerably reducing the optimization time. While convex optimization relies on analytical modeling and hence cannot replace the full design space exploration, it can be very efficient for a high-level early stage of such exploration, guiding computer architects as to the initial choice of area, voltage, and frequency of the individual components of the CMP.

**ACKNOWLEDGMENT**

This research was partially funded by the Intel Collaborative Research Institute for Computational Intelligence and by Hasso-Plattner-Institut.

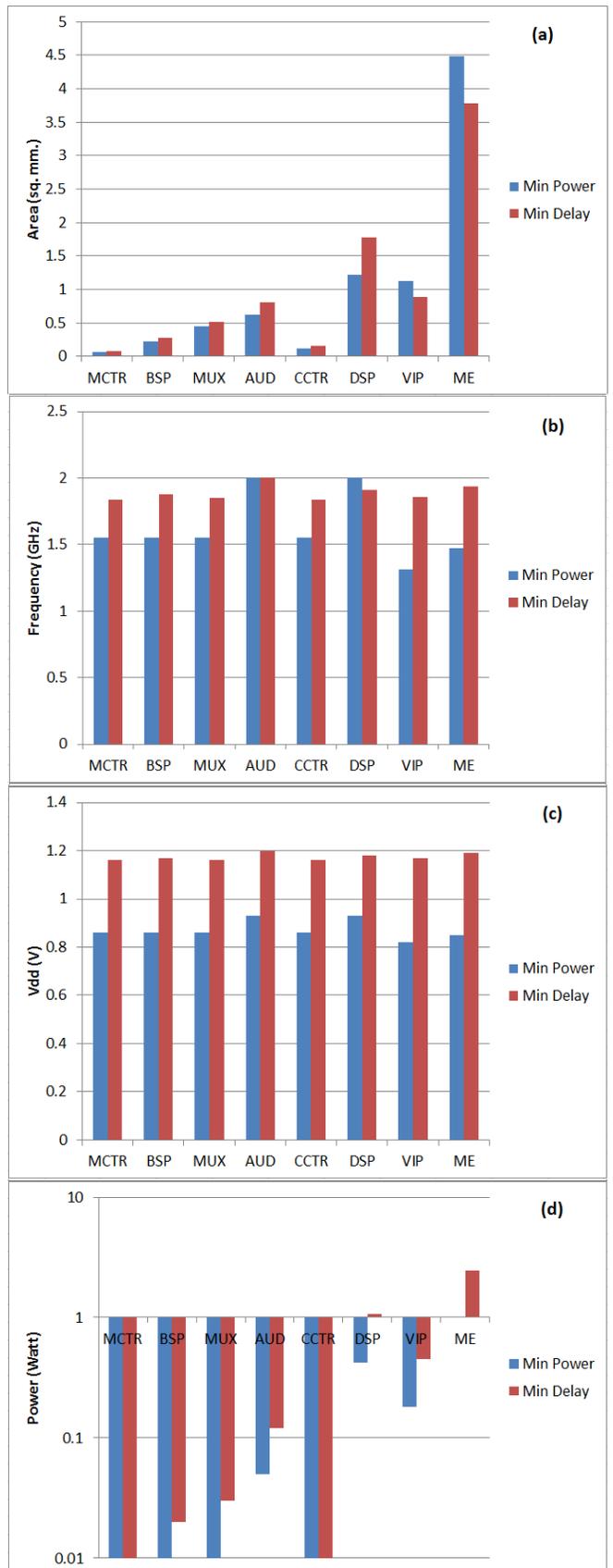

Fig. 3. Power vs. Execution Time (Delay) Optimization Results: (a) Optimal Area; (b) Optimal Frequency; (c) Optimal Supply Voltage (Vdd); (d) Optimal Power; all modeled for $\alpha=3$

# REFERENCES

[1] A. Alameldeen, "Using compression to improve chip multiprocessor performance", PhD thesis, University of Wisconsin at Madison, 2006.

[2] A. Cassidy and A. Andreou, "Beyond Amdahl Law - An objective function that links performance gains to delay and energy", IEEE Transactions on Computers, vol. 61, no. 8, pp. 1110-1126, Aug 2012.

[3] A. Elyada, R. Ginosar, U. Weiser. "Low-complexity policies for energy-performance tradeoff in chip-multi-processors." Very Large Scale Integration (VLSI) Systems, IEEE Transactions on 16.9 (2008): 1243-1248.

[4] A. Morad, T. Morad, L. Yavits, R. Ginosar, U. C. Weiser. "Generalized MultiAmdahl: Optimization of Heterogeneous Multi-Accelerator SoC," IEEE Computer Architecture Letters, 2012.

[5] Amir Morad, Leonid Yavits, Gedalia Oxman, Michael Khrapkovsky, Ofer Austerlitz, Hila Madar, "Stream multiplexer/de-multiplexer" US Patent App. 11/727,719, 2007

[6] Broadcom BCM7040 Video/Audio encoder, http://www.broadcom.com/products/Satellite/MPEG-2-Digital-Audio-Video-Encoders/BCM7040.

[7] Isci, C., Buyuktosunoglu, A., Cher, C.Y., Bose, P. and Martonosi, M., 2006, December. An analysis of efficient multi-core global power management policies: Maximizing performance for a given power budget. In Proceedings of the 39th annual IEEE/ACM international symposium on microarchitecture(pp. 347-358). IEEE Computer Society.

[8] Wentzlaff, D., Beckmann, N., Miller, J. and Agarwal, A., 2010. Core count vs cache size for manycore architectures in the cloud. MIT-CSAIL-TR-2010-008.

[9] E. Rotem, R. Ginosar, U. Weiser, A. Mendelson, "Energy Aware Race to Halt: A Down to EARtH Approach for Platform Energy Management", IEEE Computer Architecture Letters, 2012.

[10] G. Loh, "The cost of uncore in throughput-oriented many-core processors", Workshop on Architectures and Languages for Throughput Applications. 2008.

[11] J. Butts, G. Sohi. "A static power model for architects", 33rd ACM/IEEE international symposium on Microarchitecture. ACM, 2000.

[12] Huh, J., Burger, D. and Keckler, S.W., 2001. Exploring the design space of future CMPs. In Parallel Architectures and Compilation Techniques, 2001. Proceedings. 2001 International Conference on (pp. 199-210). IEEE.

[13] Skadron, K., Stan, M.R., Huang, W., Velusamy, S., Sankaranarayanan, K. and Tarjan, D., 2003. Temperature-aware microarchitecture. ACM SIGARCH Computer Architecture News, 31(2), pp.2-13.

[14] Wang, L. and Skadron, K., 2012. Dark vs. dim silicon and near-threshold computing extended results. University of Virginia Department of Computer Science Technical Report TR-2013, 1.

[15] L. Yavits, A. Morad, "Video encoding and video/audio/data multiplexing device." U.S. Patent No. 6,690,726. 10 Feb. 2004.

[16] L. Yavits, A. Morad, R. Ginosar, "Cache Hierarchy Optimization," IEEE Computer Architecture Letters, 2013

[17] L. Yavits, A. Morad, R. Ginosar, "The effect of communication and synchronization on Amdahl's law in multicore systems", Parallel Computing journal, 2013.

[18] Zhao, L., Iyer, R., Makineni, S., Moses, J., Illikkal, R. and Newell, D., 2007, February. Performance, area and bandwidth implications on large-scale CMP cache design. In Proceedings of the Workshop on Chip Multiprocessor Memory Systems and Interconnect.

[19] M. Grant, S. Boyd, Y. Ye. "CVX: Matlab software for disciplined convex programming", 2008.

[20] M. Hill, M. Marty. "Amdahl's law in the multicore era", Computer 41.7 (2008): 33-38.

[21] Hardavellas, N., Ferdman, M., Falsafi, B. and Ailamaki, A., 2011. Toward dark silicon in servers. IEEE Micro, 31(EPFL-ARTICLE-168285), pp.6-15.

[22] S. Boyd, L. Vandenberghe, "Convex optimization", Cambridge university press, 2004.

[23] Heo, S., Barr, K. and Asanović, K., 2003, August. Reducing power density through activity migration. In Low Power Electronics and Design, 2003. ISLPED'03. Proceedings of the 2003 International Symposium on (pp. 217-222). IEEE.

[24] Oh, T., Lee, H., Lee, K. and Cho, S., 2009, May. An analytical model to study optimal area breakdown between cores and caches in a chip multiprocessor. In VLSI, 2009. ISVLSI'09. IEEE Computer Society Annual Symposium on (pp. 181-186). IEEE.

[25] Yavits, L., Morad, A. and Ginosar, R., 2014", Cache hierarchy optimization", IEEE Computer Architecture Letters, 13(2), pp.69-72.

[26] Yavits, L., Morad, A. and Ginosar, R., 2014. The effect of communication and synchronization on Amdahl's law in multicore systems. Parallel Computing, 40(1), pp.1-16.

[27] Yavits. L, A. Morad, R. Ginosar, "Computer Architecture with Associative Processor Replacing Last Level Cache and SIMD Accelerator", IEEE Transactions on Computers, 2015, vol. 64, issue 2, pp 368 - 381

[28] Zidenberg, T., Keslassy, I. and Weiser, U., 2012. MultiAmdahl: How should I divide my heterogenous chip?. Computer Architecture Letters, 11(2), pp.65-68.